\documentclass[12pt]{elsart} % gallery format

\usepackage{graphicx}
\usepackage{amsmath}
\usepackage{citesort}
\usepackage{psfrag}
\usepackage{psfig}

\newcommand{\av}[1]{\left< #1 \right>}

\begin{document}

\bibliographystyle{unsrt}
%\begin{flushright}
%  \it \footnotesize
%  Dedicated to Klaus Kirchg\"assner for his 70th birthday.
%\end{flushright}
\begin{frontmatter}

\title{Modelling a Dune Field}

\author{A. R. Lima$^1$, G. Sauermann$^{1,2}$,  H. J.
  Herrmann$^{1,2}$} and
   K. Kroy$^1$ 
\address{1) PMMH, Ecole Sup\'erieure de Physique et Chimie
  Industrielles (ESPCI), 10, rue Vauquelin, 75231 Paris, Cedex 05,
  France}

\address{2) ICA-1, University of Stuttgart, Pfaffenwaldring 27, 70569
  Stuttgart, Germany}

%\date{Version: 28.05.2001 Printed: \today}

\maketitle

\begin{abstract}
We present a model to describe the collective motion of barchan dunes
in a field. Our model is able to reproduce the observation that a
typical dune stays confined within a stripe. We also obtain some of
the pattern structures which ressemble those observed from aerial photos which we do
analyse and compare with the specific field of La\^ayounne.
%  The position of dunes in a dune fields is a question that was not
%  addressed in detail up to now. In this paper we propose a simple
%  model that defines interactions between the dunes in a dune field
%  and predicts its the time evolution. Some of the effects the model
%  is able to reproduce are the correlations between the position of
%  dunes and the focusing of the whole field.
\end{abstract}

\end{frontmatter}

%%%%%%%%%%%%%%%%%%%%%%%%%%%%%%%%%%%%%%%%%%%%%%%%%
\section{Introduction}

\begin{figure}[tb]
  \begin{center}
    \includegraphics[width=0.5\textwidth]{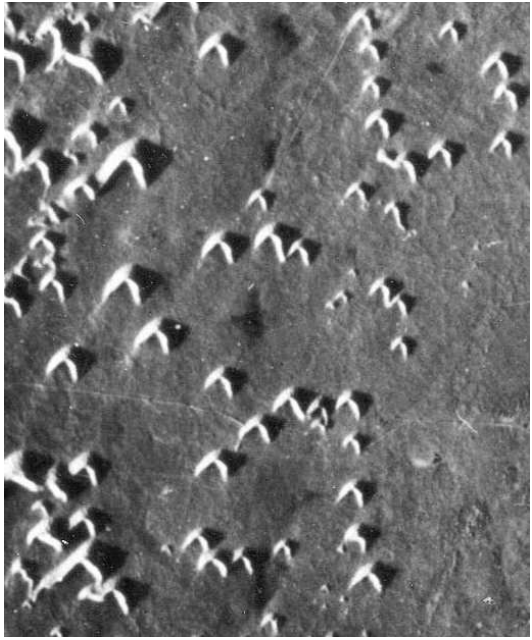}
    \caption{Aerial photograph of a barchan dune field close to
    La\^ayounne in
      Morocco. The photo clearly shows a correlation between the dunes
      in the lateral direction.}
    \label{fig:dune_corr_morocco}
  \end{center}
\end{figure}

A wide variety of dune shapes can be found in deserts, in coastal areas, on the
sea--bottom \cite{Berne89}, and even on Mars \cite{Thomas99} and over
one hundred different types of dunes have been classified by geomorphologists
and their ancient  predecessors so that some of them bear old arabic names. The shapes
depend mainly on the amount of available sand and on the change in the
direction of the wind over the year. If the wind blows steadily from
the same direction throughout the year and there is not enough sand to
cover the entire area, dunes  called {\em
  barchans} develop having the shape of a crescent. They typically appear in huge fields of several
  hundred kilometers length stretching along coastlines and driven by
  trade-winds. A picture from a section of such a field is shown in 
Fig. \ref{fig:dune_corr_morocco}.

Measurements of the relationship between height, width  
and length have been performed for instance in Peru or Morocco
\cite{Finkel59,Hastenrath67,Hastenrath87,Lettau69,Lettau78,sauermann-etal:2000}
and led to a good understanding of the barchan shape. In particular we
know that the barchans move without changing their shape with a
velocity which is inversely proportional to their height. However, the
dynamics of dunes in a dune field has not yet been studied very much. In
particular, there seems to be a contradiction between the fact that
the barchans are not all of equal height and the stability of the dune
field since when the dunes have different velocities they will
eventually bump into one another. We do not yet understand the
underlying stabilizing mechanism. Looking at the images in more
detail also other questions arise. Are the
relative positions of the dunes correlated? Do they form patterns?
When dunes are getting closer they can interact through the
non--local wind field or the sand flux between the dunes. In this work,
we propose a simple model that defines the interaction between the
dunes caused  by the inter--dune sand flux. We neglect further  effects of
the wind field. The model predicts the time evolution of the entire
field. It can reproduce the spatial correlations between
the dunes that have been observed in the field,
cf. Fig. \ref{fig:dune_corr_morocco}. Furthermore, the model shows
that the entire dune field does not broaden but stays
focused. However, the stability of the dune field depends on the sand
flux balance of a single dune, which is not well known so that we will
have to make some simplifying assumptions.

%%%%%%%%%%%%%%%%%%%%%%%%%%%%%%%%%%%%%%%%%%%%%%%%%
\section{Wind and sand flux}

Sand is transported if the wind speed exceeds a certain threshold
velocity. In equilibrium the amount of transported sand or the
flux of sand $q(u_*)$ can be expressed as a function of the shear
velocity $u_*$ \cite{Pye90}. The shear velocity $u_*$ characterizes
the turbulent boundary layer of the atmosphere and thus its well known
logarithmic velocity profile $v(z) = u_* / \kappa \ln z / z_0$, where
$v(z)$ is the wind speed at height $z$ over the ground, $\kappa
\approx 0.4$ the universal von K{\'a}rm{\'a}n constant for turbulent
flow, and $z_0$ the roughness length of the surface. For shear
velocities well above the threshold, e.g. $u_*=0.5\,$m$\,$s$^{-1}$,
the simplest sand flux relation, proposed by Bagnold \cite{Bagnold41},
should give a reasonable prediction for the sand flux $q$,
\begin{equation}
  q  = C_B \frac{\rho_a}{g} u_*^3,
  \label{eq:sand_flux}
\end{equation}
where $C_B \approx 2$ is a phenomenological constant
\cite{Bagnold41,Rasmussen91,unpub:SauermannKroy2001},
$\rho_a=1.225\,$kg$\,$m$^{-3}$ the density of air, and
$g=9.81\,$m$\,$s$^{-1}$ the acceleration due to gravity. Using the
bulk density $\rho_s = 1650\,$kg$\,$m$^{-3}$ of a typical sand we
can rewrite Eq.~(\ref{eq:sand_flux}) into a relation for the volume
flux $\Phi$,
\begin{equation}
  \Phi = \frac{q}{\rho_s}
  \label{eq:sand_flux2}
\end{equation}
For the discussion of the dune field dynamics we chose the shear
velocity over the flat ground $u_{*0}=0.5\,$m$\,$s$^{-1}$ and obtain a
volume sand flux $\Phi_0$ of 0.031$\,$m$^{2}$s$^{-1}$. Furthermore, we
assume that the sand velocity $v_s$ is proportional to the shear
velocity $u_*$ and choose a typical value of $v_s=5\,$m$\,$s$^{-1}$.

%%%%%%%%%%%%%%%%%%%%%%%%%%%%%%%%%%%%%%%%%%%%%%%%

\section{Barchan dunes}

A dune is characterized by a single scalar quantity, e.g. its width
$w$, and its position $(x,y)$ in the dune field. Further properties
such as the height $h$,
\begin{equation}
  \label{eq:height}
  h = a \, w
\end{equation}
or volume $V$, 
\begin{equation}
  \label{eq:volume}
  V = b \, w^3
\end{equation}
can be calculated using simple phenomenological formulas. Here, the
parameters $a=0.09$ and $b=0.05$ have been obtained from field
measurements \cite{sauermann-etal:2000}.

\begin{figure}[tb]
  \begin{center}
    \includegraphics[width=0.7\textwidth]{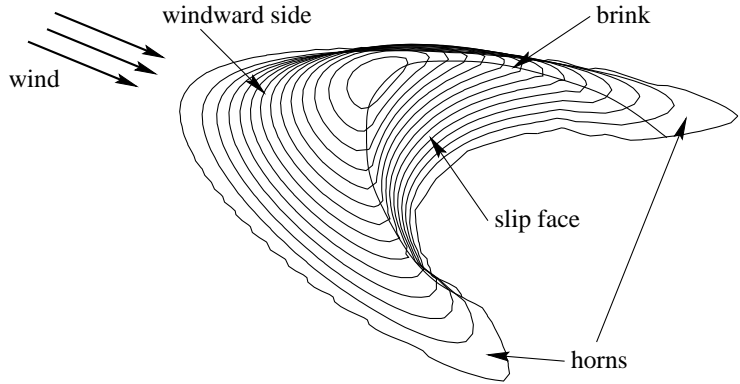}
    \caption{Sketch of a barchan dune.}
    \label{fig:dune_sketch}
  \end{center}
\end{figure}

The migration velocity $v_d$ of a dune depends on its height $h$ and the sand flux $\Phi_b$ over the brink of the dune
\cite{Bagnold41}. The latter can be calculated through 
Eq.~(\ref{eq:sand_flux}) if the shear velocity at the brink is known.
However, the dune itself disturbs the wind field such that it becomes stronger above the dune than over a flat ground. The ratio between
the shear velocity over a flat ground $u_{*0}$ and the shear velocity
at the brink of the dune is called the ``speed--up'' $n$. This speed--up
depends on the shape of the dune and mainly on the average windward
slope. A typical value for the speed--up over a sand dune is 1.4
\cite{jackson-hunt:75,hunt-leibovich-richards:88}. Finally, we can
write for the dune velocity $v_d$,
\begin{equation}
  \label{eq:dune_speed}
  v_d = \frac{\Phi_b}{h} = \frac{\Phi_0 \, n}{a w}.
\end{equation}
We want to note that Eqs.~(\ref{eq:height}) and (\ref{eq:volume})
implicitly assume scale invariant dune shapes, which is an
approximation that is only valid for large dunes, much larger than the
minimal dune size \cite{unpub:KroySauermann2001,SauermannPhD2001}.

The growth and shrinkage of a dune depends on the balance of in-- and
out--going sand fluxes. The sand flux that arrives from a upwind
location, for instance a beach or other dunes, feeds the dune with new
sand, whereas at the tip of the horns sand leaves the dune. The change
of volume $\dot V_i$ due to the in--flux $\Phi(y)$ can be calculated
by integrating over the dunes width,
\begin{equation}
  \label{eq:dvdt_in}
  \dot V_i = \int_0^w \Phi(y) \, dy.
\end{equation}
The sand that leaves the dune at the horns gives rise to a loss of
volume $\dot V_o$ that is modeled by the following expression,
\begin{equation}
  \label{eq:dvdt_out}
  \dot V_o = w \, \Phi_0 ( a_0 + a_1 w ),
\end{equation}
where $a_0$ and $a_1$ are parameters that depend on the complex
processes occurring on the dune's surface and are not known a priori.
Recent numerical calculations of the sand outflux from barchan dunes
suggest that the outflux is nearly proportional to the dunes width and
motivated the expression above \cite{SauermannPhD2001}.

For a stable dune, the gain and loss of volume must exactly compensate
$\dot V_i - \dot V_o = 0$. The average inter--dune flux or influx
$\av{\Phi} = f \, \Phi_0$ causes an average increase in volume,
\begin{equation}
  \label{eq:dvdt_av_in}
  \av{\dot V_i} = w \, f \, \Phi_0,
\end{equation}
where $f$ defines the saturation of the inter--dune flux.
Eqs.~(\ref{eq:dvdt_out}) and (\ref{eq:dvdt_av_in}) define finally an
average dune width $\left< w \right>$,
\begin{equation}
  \label{eq:in-out}
  \left< w \right> = \frac{f-a_0}{a_1}.
\end{equation}
Using the average dune width $\av{w}$ and $\gamma=a_1 \av{w} / f$ as new
parameters, instead of $a_0$ and $a_1$, we can rewrite 
Eq.~(\ref{eq:dvdt_out}),
\begin{equation}
  \label{eq:dvdt_out2}
  \dot V_o = w \, f \Phi_0 \left( 1 + \gamma \frac{w-\av{w}}
    {\av{w}} \right).
\end{equation}
Finally, the change in volume $\dot V$ of a dune can
be obtained from Eqs.~(\ref{eq:dvdt_in}) and (\ref{eq:dvdt_out2}),
\begin{equation}
  \label{eq:dvdt}
  \dot V = \int_0^w \Phi(y) dy - 
           w \, f \Phi_0 \left( 
             1 + \gamma \frac{w-\av{w}}{\av{w}}
           \right) 
\end{equation}
Typical values for $f$ are of the order of
0.1, i.e. the inter--dune flux is far from being saturated. 

%%%%%%%%%%%%%%%%%%%%%%%%%%%%%%%%%%%%%%%%%%%%%%%%%%%%%%%%%%%%

\section{The barchan field}

\begin{figure}[tb]
  \centerline{\psfig{file=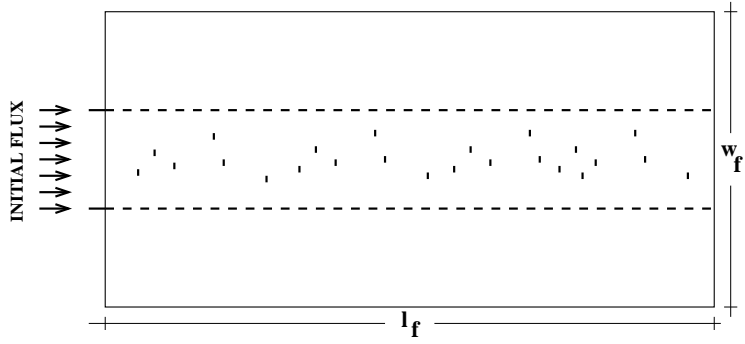,width=0.7\textwidth}}
  \caption{Sketch of a dune field.}
  \label{fig:sketchfield}
\end{figure}

A barchan field of length $l_f$ and width $w_f$ as depicted in
Figure~\ref{fig:sketchfield} is composed of many dunes that migrate in
the $x$--direction, the direction of the wind.  The center of mass
($x_i$, $y_i$) of a dune is used to localize the dune $i$ in the
field. The dynamics of the dune field is described by the temporal
change of the variables $x_i$, $y_i$, and $w_i$ which will be defined
in the following.

From Eqs.~(\ref{eq:dvdt}) and (\ref{eq:volume})
we obtain the change of the width of a dune, which depends on the
inter--dune flux,
\begin{equation}
  \label{eq:dvdt_i}
  \dot w_i = b^{-1/3} \left[ \int_{y_i-w_i/2}^{y_i+w_i/2} \Phi(y) dy - 
           w_i \, f \Phi_0 \left( 
             1 + \gamma \frac{w_i-\av{w}}{\av{w}}
           \right) \right]^{1/3}
\end{equation}

The velocity of the dune in the wind direction is given in
Eq.~(\ref{eq:dune_speed}) and depends only on the width of the dune,
\begin{equation}
  \label{eq:x_i}
  \dot x_i = v_d(w_i) = \frac{\Phi_0 n}{a w_i}
\end{equation}

An inter--dune flux that is asymmetric in the lateral direction with
respect to the center of mass $y_i$ of the dune gives rise to a change
in lateral position $y_i$,
\begin{equation}
  \label{eq:y_i}
  \dot y_i = \frac{V_i \, y_i + \int_{y_i-w_i/2}^{y_i+w_i/2}\,{\tilde y
      \, \Phi(\tilde y) \, d \tilde y}}
      {V_i + \int_{y_i-w_i/2}^{y_i+w_i/2}\,{\Phi(\tilde y)
      \, d \tilde y}} - y_i,
\end{equation}
where $V_i$ is the volume of the dune $i$.

Dunes can interact either due to the inter--dune flux or due to
coalescence. The first case is defined by the equations above and the
inter--dune sand flux $\Phi(x,y)$ defined in the next section. For the
latter case we define a characteristic distance of coalescence
$d_{\rm col}$. If two dunes become closer than this distance they are
merged to a single dune. The mass is conserved by this process and the
new position is calculated according to their centers of mass and
relative volumes.

%%%%%%%%%%%%%%%%%%%%%%%%%%%%%%%%%%%%%%%%%%%%%%%%
\subsection{The inter--dune sand flux in the field}

The flux of sand in the field, i.e. between the dunes, is a complex
problem and causes an interaction between the dunes. Dunes can act here
as sources and sinks for the sand at the windward foot and at the
horns, respectively. Furthermore, the position and width of the dunes
changes in time and thus their influence on the inter--dune sand flux.
We propose the following model for the inter--dune flux: First, the
sand is convected with the constant velocity $v_s$ in the wind
direction. Secondly, a lateral movement is caused  by
diffusion and obeys a diffusion equation,
\begin{equation}
\frac{\partial \Phi}{\partial t} - D \frac{\partial^2
  \Phi}{\partial y^2} = 0,
\label{eq:diff}
\end{equation}
where the time scale of the diffusion process is connected to the
transport in wind direction by the sand velocity $v_s$. Hence, we
start solving the diffusion equation at the influx--border $x=0$ of
the field, where the sand flux is spatially constant. When integrating
the solution forward in time by displacing its $x$--position we obtain the stationary solution of the
two--dimensional inter--dune flux field $\Phi(x,y)$,
\begin{equation}
  v_s \frac{\partial \Phi}{\partial x} - D \frac{\partial^2
  \Phi}{\partial y^2} = 0.
\label{eq:diff_x}
\end{equation}

For each dune in the field, the flux $\Phi_i$ is set to zero along the
entire width of the dune (from $y_i-w_i/2$ to $y_i+w_i/2$). Then, at
the horns (i.w. the positions $y_i-w_i/2$ and $y_i+w_i/2$) the outflux is imposed as two
$\delta$-functions. Finally, the new dune mass is calculated by the
in-- and out--flux balance, Eq.~(\ref{eq:dvdt_i}).

%%%%%%%%%%%%%%%%%%%%%%%%%%%%%%%%%%%%%%%%%%%%%%%%
\subsection{Initial configuration and Boundary conditions}

The field is initialized with dunes at random positions until a
certain density $\rho_d$ is reached, which is a parameter of the
model. The dune width is chosen between a minimum and a maximum width,
$w_{min} \le w \le w_{max}$, having an average width
$\av{w}=(w_{min}+w_{max})/2$. From the initial density $\rho_d$ we can
define the rate $r$ with which new dunes enter the field at the in--flux
boundary,
\begin{equation}
  r = W \, \rho_d \, v(\av{w}),
\end{equation}
where $v(\av{w})$ is the dune velocity according to
Eq.~(\ref{eq:dune_speed}) and $W$ is the width of the dune field,
which is chosen to be only one third of the total simulation area, cf.
Fig~\ref{fig:sketchfield} where $w_{f} = 3W$. The width of the dunes that enter with the
rate $r$ at the influx boundary is chosen randomly as in the initial
configuration. If a dune crosses the outflow boundary (at the right
side of the field) it is removed from the simulation.

The inter--dune flux of sand $\Phi$ is chosen to be constant between
$w_f/3 < y < 2 w_f/3$ and zero outside. The constant influx is chosen
to be $\Phi= f \Phi_0$ and thus corresponds to an inter--dune flux of a
dune field with average dune width $\av{w}$, cf. Eq.
(\ref{eq:dvdt_av_in}). In the direction perpendicular to the wind
($y$--direction) periodic boundary conditions are used.

%%%%%%%%%%%%%%%%%%%%%%%%%%%%%%%%%%
\section{Numerical Scheme}

The numerical scheme can be divided into several steps. First, the
dune field is randomly initialized with dunes as discussed above.
Then, an iterative calculation, forward in time, of the following
steps is performed:
\begin{itemize}
\item New dunes are created with rate $r$ at the left boundary, in the
  region $w_f/3 < y < 2 w_f/3$.
\item The new $x$--position of the dunes is calculated according to
  Eq.~(\ref{eq:x_i}).
\item If two dunes are closer than the coalescence distance $d_{\rm
    col}=50\,$m they are merged.
\item The inter--dune sand flux is calculated from left to right on a
  discrete one dimensional grid in $y$--direction according to
  Eq.~(\ref{eq:diff_x}). The iteration is started at $x=0$ with a
  constant value of $\Phi(y)=f \Phi_0$ for $w_f/3 < y < 2 w_f/3$ and zero
  outside. When the calculation forward in the wind direction reaches
  a dune, the flux is set to zero along the dune width and two
  discrete $\delta$--peaks are set at the horns of the dune
  as described above.  Furthermore, the change in volume of the dune
  and its lateral movement according to Eqs.~(\ref{eq:dvdt_i}) and
  (\ref{eq:y_i}) are calculated.
\end{itemize}
The time step between the iteration cycles above has been chosen to be
$\Delta T = 1/12$ year.

%parameters:
%$f=0.15$, $\gamma=\text{unknown}$, $D=\text{unknown}$

%%%%%%%%%%%%%%%%%%%%%%%%%%%%%%%%%%%%%%%%%%%%%%%%%%%%%%%%%%%%
\section{Results}

We are mainly interested in measuring how the dunes are distributed in
the field, possible correlation between the positions of the dunes and
understanding the confinement of the field on a stripe.  We did simulations of our model for
a field having a length of $50$ km  and a width of $12$ km. The dunes enter 
the field only in the region between km $4$ and km $8$. The initial
density of dunes in the field was chosen to be  $\rho_f=10$
dunes/km$^2$. We set the saturation of the flux to be $f = 0.15$ which
corresponds to a flux of $\phi = 0.0046$ m$^2$/sec.

\begin{figure}[!h]
  \centerline{\psfig{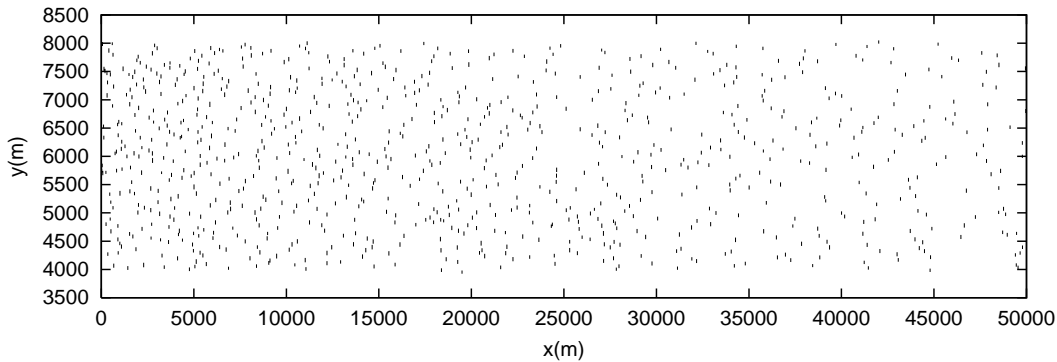}}
  \caption{Snapshot of the simulation of the field.}
  \label{fig:image}
\end{figure}

\begin{figure}[!h]
  \centerline{\psfig{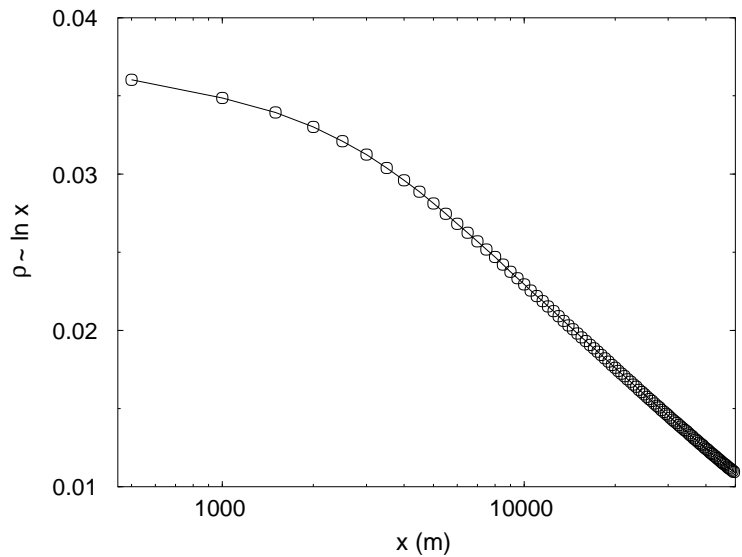}}
  \caption{Time averaged density of the dunes along the field.}
\label{fig:rhox}  
%\label{fig:evol}
\end{figure}

In fig.~\ref{fig:image} we see a typical simulation of the dune
field. It stays confined inside a stripe. The dune density slowly
decreases in direction of the wind. This can be seen quantitatively in
Fig.\ref{fig:rhox} where the density averaged over the $y$ direction
is plotted over the $x$-axis ($x$ and $y$ directions are defined in
fig.~\ref{fig:image}) in a semilogarithmic plot. We recognize for large
$x$ a logarithmic decay of the form $\rho \propto ln~x$. 

\begin{figure}[!h]
  \centerline{\psfig{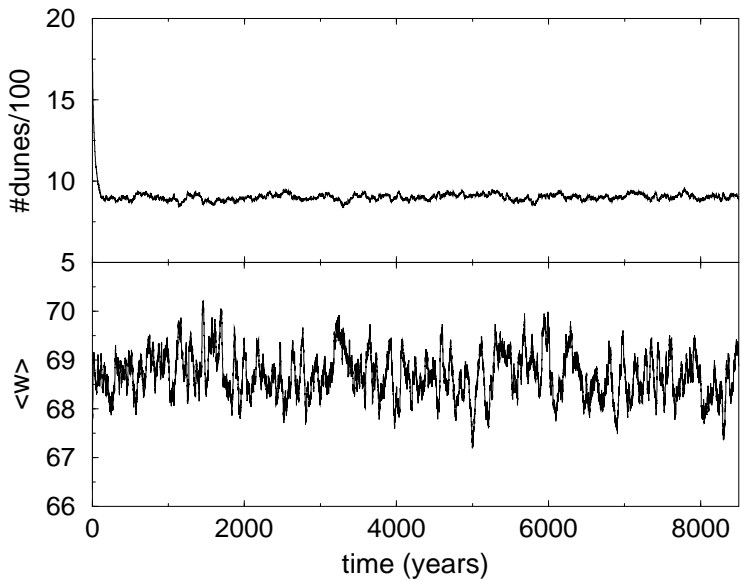}}
  \caption{Number and average size of dunes in the field}
  \label{fig:evol}
\end{figure}

The situation seen in fig.~\ref{fig:image} is a stationary state. As seen in
fig.~\ref{fig:evol} the average size of the dunes fluctuates around a
value that is constant in time and except for a short transient at
early times also the number of dunes is essentially constant in time.  

\subsection{The confinement of the field}

\begin{figure}[!h]
  \centerline{\psfig{bbllx=95pt,bblly=70pt,bburx=560pt,bbury=725pt,angle=270,file=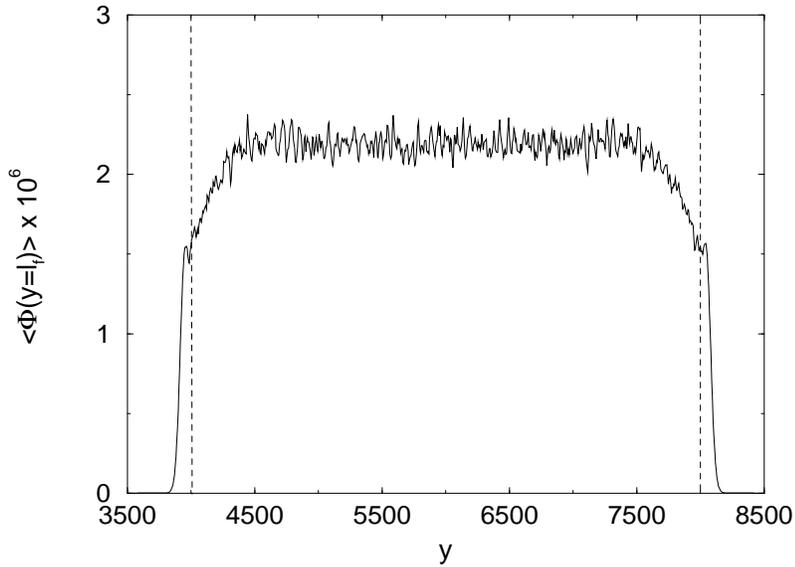,width=0.75\textwidth}}
  \caption{Averaged flux at the end of the field.}
  \label{fig:fluxy}
\end{figure}

It is not clear up to now why real dune fields are typically confined
on a stripe 
all the time. In our simple model, despite the fact that the diffusion
of the sand tends to spread the sand to the borders, this behavior
appears by itself. As seen in fig.~\ref{fig:fluxy}, the sand flux decreases outside
the central region of the field. Hence, dunes at the borders of that
region receive less sand. The ones close to the center of the
field  receive more sand. This flux gradient acts like a
``force'' which pulls the dunes into the region where there is more
flux, i.e. the central region where the other dunes are.

\begin{figure}[!h]
  \centerline{\psfig{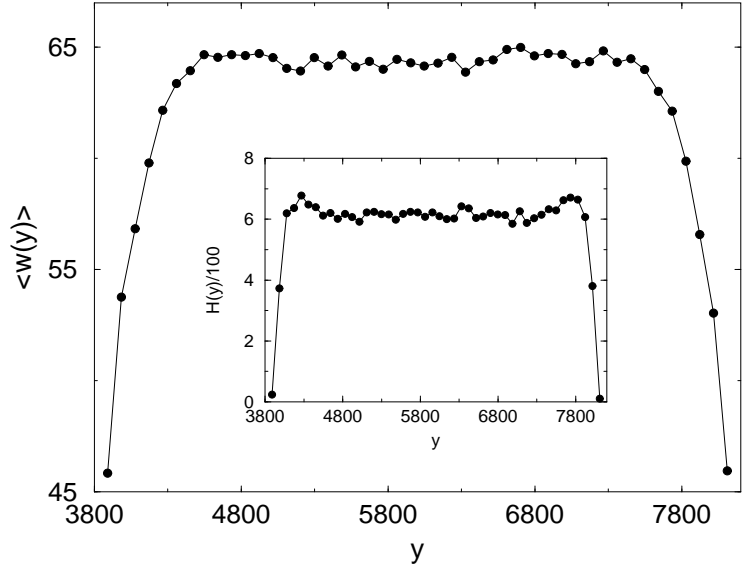}}
  \caption{Average size and number of dunes (inset) which leave a
  field of $50$ km. We can see that the field stays confined and that the dunes
    outside of the central regions tend to shrink.}
  \label{fig:disty}
\end{figure}

A way to see this effect is the distribution of dunes and their mean size at the end of
the field. This is shown in figure
(\ref{fig:disty}). In the inset we plotted the number of dunes that
leave the field at the right boundary. Near to
the center of the field the dunes have approximately the same size,
however the ones  near to the borders are
smaller, since they receive less sand. Since the dunes in the borders are
smaller, they are faster and this explains why we have more dunes
leaving the field close to the borders than in the center.

\subsection{Spatial correlations}

Looking at fig.~\ref{fig:dune_corr_morocco} we can observe that the
dunes tend to form lines, the horn of one dune pointing to the center
of the back of another one. These patterns are apparently due to the
coupling through the inter-dune sand flux and we will show how they
are reproduced by our model. In order to quantitatively measure
these structures, we will introduce two types of correlation functions
which we will monitor as well in our model as also directly from a photograph
of a real dune field in order to compare the patterns formed by our model to the observed ones.

Let us define the longitudinal correlation as 
\begin{equation}
C_l(d) = \sum_{\rm i} \sum_{j>i} {\delta (d-|x_i-x_j|)}
\end{equation} 

This function measures the correlation of dune positions in the wind
direction.

\begin{figure}[!h]
  \centerline{\psfig{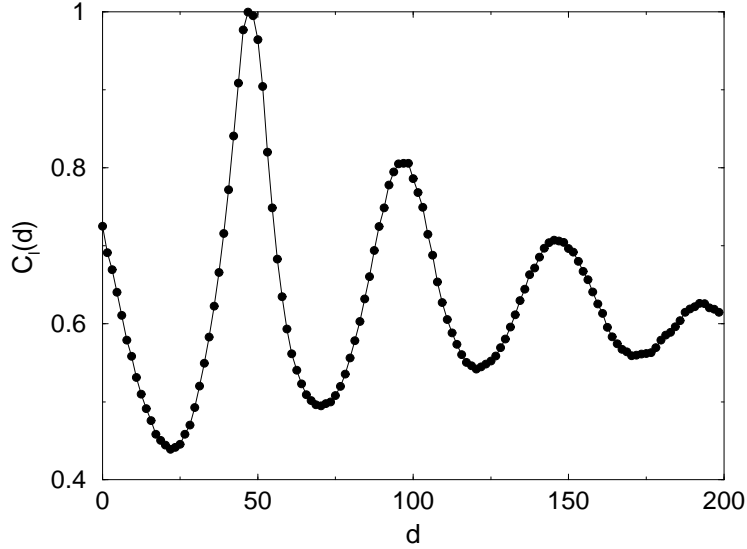}}
  \caption{Correlations in the wind direction obtained from our model
  after averaging over time.}
  \label{fig:cld}
\end{figure}

In fig.~\ref{fig:cld} we present the results for our model. To get
better statistics we take the average  over several
thousand time-steps. We
observe very pronounced peaks  at regular distances of about 50 m showing a
characteristic distance between the dunes in wind direction.

\begin{figure}[!h]
  \centerline{\psfig{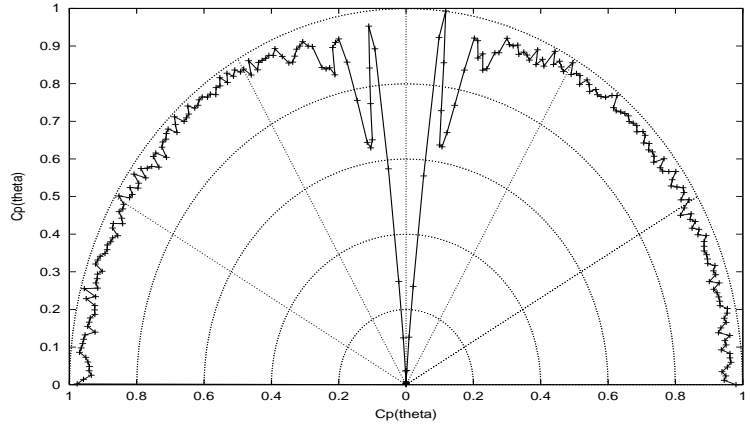}}
  \caption{Polar plot of the angular correlation $C_{p} (\theta)$  between the dunes.}
  \label{fig:cpt}
\end{figure}

It is also interesting to calculate the angular correlations in the positions
of the dunes which can be quantified by the angular correlation function

\begin{equation}
C_p(\theta) = \sum_{\rm i} \sum_{j>i} {\delta \left(\theta-arctan\left(\frac{y_i-y_j}{x_i-x_j}\right)\right)}
\end{equation} 

where $i$ and $j$ are only chosen within the central part of the
stripe (about one third of all dunes) in order to avoid boundary
effects and for which the distance $(x_{i} - x_{j})^{2} + (y_{i} +
y_{j})^{2} < (5 \rm{km})^{2}$. In fig.~\ref{fig:cpt} we
see our results for $C_{p} (\theta)$ as obtained after averaging over
all dunes in the central part of  the field and several thousand time-steps.

We see that while most directions have no particular  structure in the
forward direction there are two favorite directions symmetrical around
the exact wind direction ($x$-axis) of high correlation. In the wind
direction itself the correlation is particularly low. The selected
direction corresponds to the arrangements where the dune in front is
just shifted by half a dune width, so that the horns directly point to
the center of the dune in front. Such configurations can be recognized
with the naked eye on the photo of fig.~\ref{fig:dune_corr_morocco}.

%%%%%%%%%%%%%%%%%%%%%%%%%%%%%%%%%%%%%%%%%%%%%%%%%%%%%%%%%%%%
\section{Results from the experiments}

\begin{figure}[!h]
  \centerline{\psfig{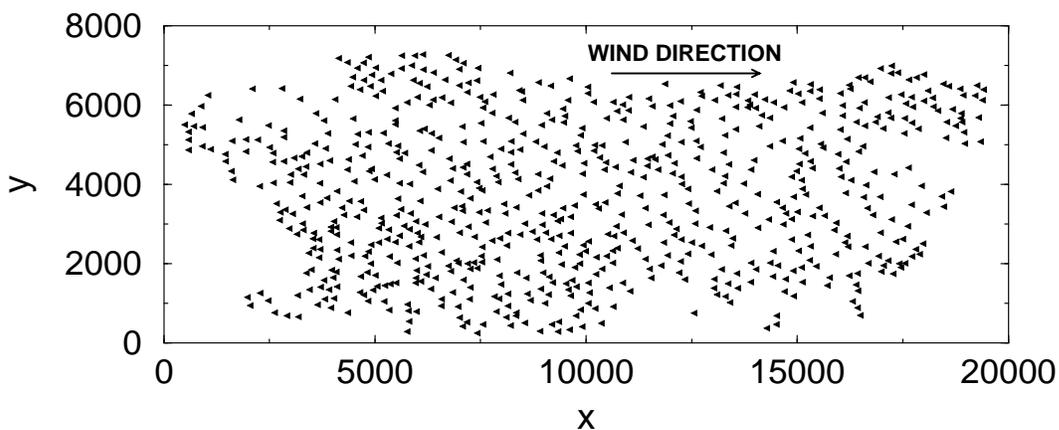}}
  \caption{Position of the dunes in the Barchan field of La\^ayounne obtained from an aerial
    photo.}
%  \label{fig:cpt}
\label{fig:real}
\end{figure}

\begin{figure}[!h]
  \centerline{\psfig{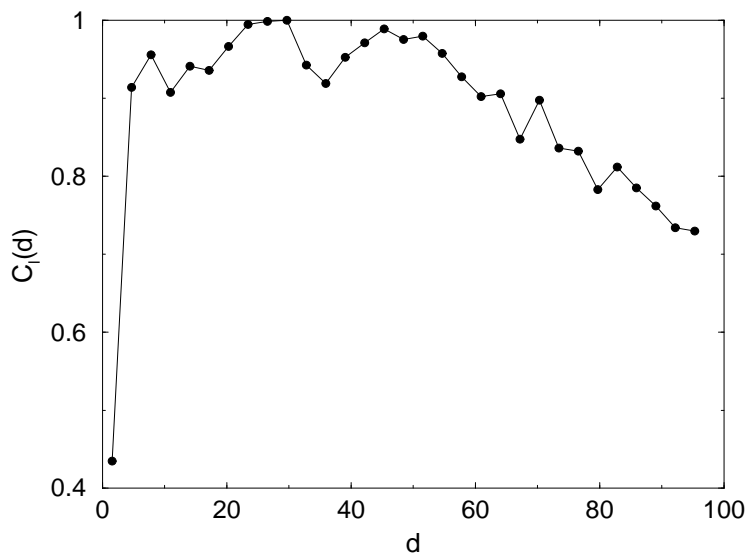}}
  \caption{Correlations in the wind direction obtained from the
  positions of the dunes in fig.~\ref{fig:real}.}
  \label{fig:cld-real}
\end{figure}

\begin{figure}[!h]
  \centerline{\psfig{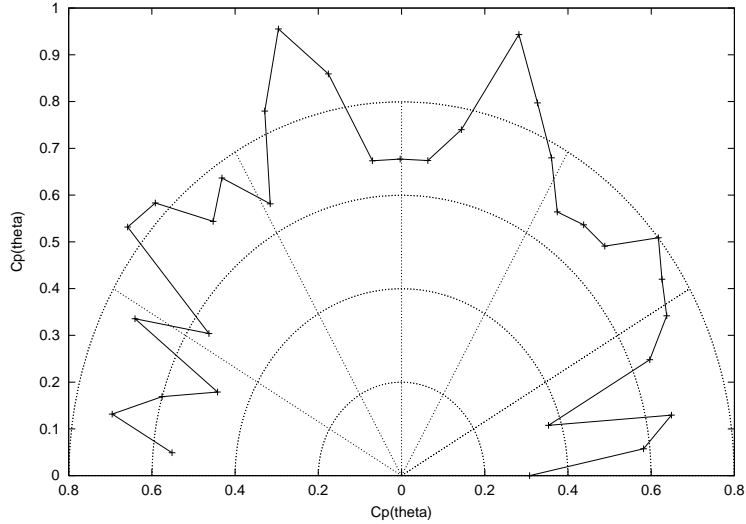}}
  \caption{Polar plot of the angular correlations $C_{p}(\theta)$ of the dune field of
  La\^ayounne.}
  \label{fig:cpt-real}
\end{figure}

We analysed  aerial photos of the dune field of La\^ayounne in
Southern Morocco and obtained the position of 734
dunes as shown in fig.~\ref{fig:real}. In  figs.~\ref{fig:cld-real} and
\ref{fig:cpt-real}  we show our results for both correlation functions
corresponding to the correlation functions obtained for the model in
figs.~\ref{fig:cld} and \ref{fig:cpt}. Comparing figs.~\ref{fig:cld}  
and \ref{fig:cld-real} one sees that the measured longitudinal
correlation does not show much periodicity or characteristic
distance. We believe that this lack of structure is rather due to
insufficient statistics. After all the model gives the possibility to
the average over time while the measurements are obtained from a
single photo.

Concerning the polar correlations, fig.~\ref{fig:cpt-real} clearly
shows the same two peaks pointing in both directions with an angle of
$\theta$ from the wind direction. But while in the model $\theta$ was
about 8$^0$, it is about 15$^0$ for the real data.

%%%%%%%%%%%%%%%%%%%%%%%%%%%%%%%%%%%%%%%%%%%%%%%%%%%%%%%%%%%%
\section{Conclusions}

In this paper we introduced a  simple model for a field of Barchan
dunes. The dunes are characterized by their width and position in the
field. The model includes a diffusion of the sand. Assuming a simple law for the inter-dune sand flux we could show
that the simulated field is stably confined and presents the same kind
of spatial correlations as real fields.

An extension of this work could be to introduce variations in the wind
direction. This could help to explain more complex patterns observed
in deserts.

%%%%%%%%%%%%%%%%%%%%%%%%%%%%%%%%%%%%%%%%%%%%%%%%%%%%%%%%%%%%
\section{Acknowledgement}

ARL acknowledges a fellowship from CNPq (Brazilian Agency) and Gerd
Sauermann a fellowship from DAAD.

\bibliography{journals,dune,books,unpublished}
\end{document}